\documentstyle[12pt]{article}
\newcommand{\be}{\begin{equation}}
\newcommand{\ee}{\end{equation}}
\newcommand{\bea}{\begin{eqnarray*}}
\newcommand{\eea}{\end{eqnarray*}}
\newcommand{\ba}{\begin{eqnarray}}
\newcommand{\ea}{\end{eqnarray}}

\begin{document}

\begin{titlepage}
\begin{flushright}
{\large \bf UCL-IPT-97-09}
\end{flushright}
\vskip 1.9cm
\begin{center}

{\Large \bf Field Theory approach to $K^0-\overline{K^0}$ and
$B^0-\overline{B^0}$ systems}
\vskip 1cm

{\large M. Beuthe$^{a,}\ $\footnote{ IISN researcher under contract
4.4509.66}, G. L\'{o}pez Castro$^{a,\ b}$ and J. Pestieau$^a$}\\

{\it $^a$ Institut de Physique Th\'eorique, Universit\'e catholique de
Louvain,}\\
{\it  B-1348 Louvain-la-Neuve, Belgium}

{\it $^b$ Departamento de F\'\i sica, Cinvestav del IPN, Apartado Postal}\\
{\it 14-740, C.P. 07000 M\'exico,D.F. M\'exico}

\end{center}

\vskip 1.6cm

\begin{abstract}
Quantum field theory provides a consistent framework to deal with 
unstable particles. We present here an approach based on field theory to 
describe the production and decay of unstable $K^0-\overline{K^0}$ and  
$B^0-\overline{B^0}$ mixed systems. The formalism is applied to compute 
the time evolution amplitudes of $K^0$ and $\overline{K^0}$ studied in 
DAPHNE and CPLEAR experiments. We also introduce a new set of parameters 
that describe CP violation in $ K \rightarrow \pi\pi$ decays without 
recourse to isospin decomposition of the decay amplitudes.

\end{abstract}

\


\end{titlepage}%

\medskip

\newpage
\section{Introduction}

  Neutral strange and beauty pseudoscalar mesons, $ K^0\overline{K^0}$
and $ B^0\overline{B^0}$, are systems of two unstable mixed states of
special interest for the study of weak interactions. They are
particularly suited to study the phenomena of CP violation together with
the oscillations in their time-dependent decay probabilities 
\cite{kabir,pdg}.

 The traditional description of unstable neutral kaons is based on the
Wigner-Weisskopf (WW) formalism \cite{ww}. In this approach, the time
evolution
of decaying states is governed by a Schr\"odinger-like equation based on
a {\it non-hermitian} hamiltonian \cite{sachs} that allows particle decays.
As a result,  the diagonalizing  transformations, in general, are not
unitary, the corresponding eigenstates are not orthogonal and the normalization
cannot be done without ambiguities.

   Besides these unsatisfactory features of the WW formalism, one
faces other difficulties. Projected factories of $K$ and $B$ mesons
\cite{daphne,slac} are
expected to measure the CP violation and oscillation parameters to a
higher accuracy than present experiments. While it is not clear whether
the approximations involved in the WW formalism are valid for both the
$K$ and $B$ systems, a consistent scheme is certainly required to compute
these observables to a high degree of accuracy.

   In this paper we adopt the view that the quantum mechanical behavior
of a complete process involving the production and decay of unstable
states can only be consistently described in the framework of quantum
field theory \cite{veltman}. In QFT, the S-matrix amplitude becomes the 
basic object 
that describes the properties of a physical process among particles. This
amplitude is taken between {\it in}- and {\it out}- asymptotic states
which are defined as non-interacting states (stable particles) existing far
away   the interaction region. Therefore, as a general rule, unstable
particles cannot be considered as asymptotic states.

 Under these conditions, unstable particles appear only as intermediate
states to which we associate Green functions (propagators) to describe
the propagation amplitudes from their production to their decay spacetime
locations. The form of these propagators, which is consistent with special
relativity and causality, determines the time evolution of the decay
probabilities. Since Lorentz covariance is implicit to the field theory
approach, neither boost transformations nor the choice of a specific
frame \cite{gw} are required to define the time parameter in the amplitude.

  In this paper we will also address some questions related to the usual
treatment of CP violating parameters.
  As is well known, the  $K^0\overline{K^0}$ (and $ B^0\overline{B^0}$)
system requires two parameters to account for CP violation in the
propagation ({\it indirect}) and decay ({\it direct}) of neutral kaons,
usually related to two theoretically defined complex parameters called
respectively
$\varepsilon$ and $\varepsilon'$ \cite{mrr}. On the one hand the
description based on the WW formalism is not valid beyond order
$\varepsilon$
because of the aforementioned difficulty in the normalization of
non-orthogonal states. Since $\varepsilon' \sim {\cal O}(\varepsilon^2)$
for the $ K^0\overline{K^0}$ system, it becomes necessary to establish a
correct formalism \cite{sanda} to account consistently for terms of order
$\varepsilon^2$. This is all the more needed because the usual
approximations for neutral kaons in the WW formalism might fail in the
case of $B$ mesons \cite{gibson}.

On the other hand, the reduction of observable CP violating parameters in
the $
K^0\overline{K^0}$ system to only two theoretical parameters
$\varepsilon$ and $\varepsilon'$ cannot be done without assuming isospin
symmetry and the factorization of strong rescattering effects \cite{mrr}.
These assumptions are rather strong in view of the smallness
of direct CP violating effects \cite{epsip,wiwo}. In this paper, we give up
the isospin decomposition of the amplitudes, parametrizing CP violation in
terms of three parameters. The first, $\hat \epsilon$, describes the mixing 
of CP eigenstates $K_1-K_2$
(indirect CP violation), while the two other, $\chi_{+-}$ and $\chi_{00}$,
account for the CP violating 2$\pi$ decays of $K_2$ (direct CP violation) in 
our approach.

  The paper is organized as follows. In section 2 we discuss the
diagonalization of mixed propagators in momentum space for the system of
unstable neutral pseudoscalar $K$ and $B$ mesons.  In section 3 we
focus on the space-time representations of these propagators. Section 4
is devoted to the applications of our formalism to compute the
time-dependent distributions of neutral kaon decays as adapted to CPLEAR
and DAPHNE experiments. Our conclusions are presented in section 5.

\section{Unstable particle propagator in momentum space}

  As previously discussed, the propagator is the basic object in the
S-matrix amplitude that describes the propagation of an unstable state
from its production at space-time point $x$ to its decay at point
$x'$. In this section we study the momentum space representation
of the propagator for the neutral kaon system, which will be needed to
compute the S-matrix amplitudes.

The description of the evolution of an unstable particle requires non
perturbative information to be introduced in the bare propagator. The full
propagator is obtained from a Dyson summation of self-energy graphs.
Since the weak interaction couples the flavor states $K^0$ and
$\overline{K^0}$, the renormalized propagator for these two unstable particles
is a non diagonal 2 $\times$ 2 matrix \cite{baulieu}.
By imposing the CPT symmetry, we can parametrize the inverse propagator for
unstable kaons of four-momentum $p$ as follows
\be
i \; D^{-1}(p^2) = \left( \begin{array}{ccc}
d & a+b \\
a-b & d
\end{array} \right)
\ee
where
$$
d \equiv p^2 - m^2_0 - i \; {\cal I} m \Pi_{00}(p^2),
\eqno{(2a)}
$$
$$
a + b\equiv -\Pi_{0 \overline 0}(p^2), \ \ \ \ \ \ \ \ \ \
\eqno{(2b)}
$$
$$
a - b \equiv -\Pi_{\overline 0 0}(p^2), \ \ \ \ \ \ \ \ \
\eqno{(2c)}
$$
where $m_0$ is the renormalized mass and $-i \: \Pi_{\alpha \beta}(p^2)$
with $\alpha,\beta = 0,\overline 0$ are the renormalized
complex self-energies of the neutral kaon system in an obvious notation.
Remark that the non diagonal
terms depend on the phase convention chosen for the kaons.

We define the CP eigenbasis as
\setcounter{equation}{2}
\be
\left( \begin{array}{c}
K_1 \\ K_2
\end{array}
\right) \ \equiv \ \frac{1}{\sqrt{2}} \left( \begin{array}{ccc}
1 & 1 \\
1 & -1
\end{array} \right) \ \left( \begin{array}{c}
K^0 \\ \overline{K^0}
\end{array}
\right) \ \equiv \
S \left( \begin{array}{c}
K^0 \\ \overline{K^0}
\end{array}
\right)
\ee
where $S = S^{-1}$ and CP$|K^0 \rangle = | \overline{K^0} \rangle$. The
corresponding inverse propagator is
\be
i \; \overline D^{-1} (p^2) \ \equiv \ S \; i D^{-1} (p^2) S^{-1} \ = \
\left(\begin{array}{ccc} d+a & -b \\
b & d-a
\end{array}
\right).
\ee

Now, if we introduce the complex parameter $\hat \varepsilon$ as
\be
\frac{\hat \varepsilon}{1+\hat \varepsilon^2} \ \equiv \ \frac{b}{2a} \ \ ,
\ee
we can diagonalize the inverse propagator as follows
 \be
i \; \overline D^{-1} (p^2) \ = \
\frac{1}{1-\hat \varepsilon^2}
\left(\begin{array}{ccc}
1 & \hat \varepsilon \\
\hat \varepsilon & 1
\end{array}
\right) \
\left(\begin{array}{ccc}
d+a \ \frac{1-\hat \varepsilon^2}{1+\hat \varepsilon^2} & 0 \\
0 & d-a \ \frac{1-\hat \varepsilon^2}{1+\hat \varepsilon^2}
\end{array}
\right) \
\left(\begin{array}{ccc}
1 & - \hat \varepsilon \\
-\hat \varepsilon & 1
\end{array}
\right).
\ee

Therefore, the physical basis of neutral kaons consists of two states
$K_{L,S}$ of definite masses $m_{L,S}$ and decay widths $\Gamma_{L,S}$,
such that
$$
d_S \equiv p^2-m^2_S + i m_S \Gamma_S = d + a
\frac{1-\hat \varepsilon^2}{1+\hat \varepsilon^2}
\eqno{(7a)}
$$
$$
d_L \equiv p^2-m^2_L + i m_L \Gamma_L = d - a
\frac{1-\hat \varepsilon^2}{1+\hat \varepsilon^2} \ ,
\eqno{(7b)}
$$
The constant width approximation will be justified in section 3. Consistency
between Eqs.(2a) and (7) demands to approximate ${\cal I} m \Pi_{00}(p^2)$ by
a constant term $-m_0 \Gamma_0$.
The propagator $\overline{ D} (p^2)$ now reads
\setcounter{equation}{7}
\be
-i \; \overline{ D} (p^2) = \frac{1}{1-\hat \varepsilon^2}
\left(\begin{array}{ccc}
1 & \hat \varepsilon \\
\hat \varepsilon & 1
\end{array}
\right)
\
\left(\begin{array}{ccc}
d^{-1}_S   & 0 \\
0 & d^{-1}_L
\end{array}
\right)
\
\left(\begin{array}{ccc}
1 & - \hat \varepsilon \\
- \hat \varepsilon & 1
\end{array}
\right).
\ee

  As already anticipated, the diagonalization of the non-hermitian matrix
given in Eq.(4) involves a non-unitary matrix. In order to provide a link
with the usual formalism, we can obtain a proper orthogonal and normalized
physical basis if we define independent {\it ket} (in-) and {\it bra} (out-)
states, respectively, as left-hand and right-hand eigenvectors of the inverse
propagator \cite{sachs} :
\be
\left( \begin{array}{c}
|K_S \rangle \\ |K_L \rangle
\end{array}
\right) \ \equiv \ \frac{1}{\sqrt{1-\hat \varepsilon^2}} \left(
\begin{array}{cc} 1 & \hat \varepsilon \\
\hat \varepsilon & 1
\end{array} \right) \ \left( \begin{array}{c}
|K_1 \rangle \\ |K_2 \rangle
\end{array}
\right)
\ee
and
\be
\left( \begin{array}{c}
\langle K_S| \\ \langle K_L|
\end{array}
\right) \ \equiv \
\frac{1}{\sqrt{1-\hat \varepsilon^2}}
\left(\begin{array}{cc} 1 & -\hat \varepsilon \\
-\hat \varepsilon & 1
\end{array} \right)
\left( \begin{array}{c}\langle K_1| \\ \langle K_2 | \end{array}
\right) \ .
\ee
Notice that for an arbitrary $\hat \varepsilon$ bra states do not 
correspond to hermitian conjugate of ket states.

The quantities $m_{S,L}, \Gamma_{S,L}$ can be measured
experimentally, while the parameters $a,b,m_0$ and $\Gamma_0$ can be in
principle computed from the theory. The relationships between these
two sets of parameters are
$$
\begin{minipage}{135mm}
$$
a = \frac{1}{2} \left( \frac{1+\hat
\varepsilon^2}{1-\hat \varepsilon^2}\right) \{m^2_L - m^2_S - i (m_L
\Gamma_L - m_S \Gamma_S)\}\ , \ \ \ \ \ \ \  \ \ \ \ \ \ \  \ \ \ \ \ \
\  \ \ \ \ \ \ \
\eqno{(11a)}
$$
$$
m^2_0 - i m_0 \Gamma_0 = \frac{1}{2} \{ m^2_L + m^2_S - i (m_L \Gamma_L + m_S
\Gamma_S ) \}\ ,  \ \ \ \ \ \ \  \ \ \ \ \ \ \  \ \ \ \ \ \ \  \ \ \ \ \ \ \
\eqno{(11b)}
$$
$$
b = \frac{\hat \varepsilon}{1-\hat \varepsilon^2} \{m^2_L - m^2_S - i (m_L
\Gamma_L - m_S \Gamma_S) \}.  \hspace*{50mm}
\eqno{(11c)}
$$
\end{minipage}
$$

\section{Space-time evolution of resonance propagators}

In this section we are interested in the time dependent properties of the
propagation of unstable particles for the purposes of studying CP
violation and the time oscillations in the kaon system. We shall
therefore focus on the properties of the unstable state propagator in
configuration space.

 Let us first consider the propagator for a stable spin zero particle :
\setcounter{equation}{11}
\be
\Delta_F (x'-x) \ = \ i
\int \ \frac{d^4p}{(2\pi)^4} \ \frac{e^{-i p.(x'-x)}}{p^2-m^2+i\varepsilon}.
\ee
The time dependence will become manifest in the amplitude if we put this
expression into another form showing a separate time evolution for the
particle and the antiparticle. A contour integration in the complex $p^0$
plane gives
\ba
\Delta_F (x'-x) &=&
\int \ \frac{d^3 p}{(2\pi)^3} \
\frac{e^{i \vec p .(\vec x' - \vec x)} e^{-i E(t'-t)}}{2E} \ \theta (t'-t)
\nonumber \\
& &
+ \; \int \ \frac{d^3 p}{(2\pi)^3} \
\frac{e^{-i \vec p.(\vec x' - \vec x)} e^{i E(t'-t)}}{2E} \ \theta
(t-t')
\ea
with $E \equiv \sqrt{\vec p^{\ 2} + m^2}$.

Depending of the specific process, the first (second) term in Eq. (13)
will survive in the time-dependent amplitude and will
describe a particle (antiparticle) propagating forward in time.

Let us now consider the propagator of a spin zero resonance. The
Dyson summation of self-energy graphs leads to the following
renormalized propagator in momentum space representation :
\be
\frac{i}{p^2-m^2-i \; {\cal I} m \Pi(p^2)}\ ,
\ee
where $-i \: \Pi(p^2) $ is the renormalized self-energy whose absorptive
part vanishes
under a threshold $p^2_{th}$ in the case of only one decay channel.

  In order to justify the constant width approximation used in Eqs. (7), 
 let us consider the 2$\pi$ contribution to the kaon self-energy. A direct
computation
of ${\cal I} m \Pi(p^2)$, for a kaon of squared four-momentum $s$, gives
\be
 {\cal I} m \Pi(s) = \frac{ -(g^2_1+g^2_2/2) }{8 \pi} \left 
(\frac{s-s_{th}}{s} \right)^{1/2} \theta(s-s_{th})
\ee
where $s_{th} =4m_{\pi}^2$ (with the approximation $m_{\pi^+} = m_{\pi^0}$)
and where
$g_1$, $g_2$ are the effective couplings for $K^0 \to \pi^+\pi^-,\ 
\pi^0\pi^0$ respectively.

Remarking that cutting rules give ${\cal I} m \Pi(s=m^2) = - m \Gamma$
where $\Gamma$
is the particle width in the center-of-mass frame and $m$ the kaon mass, we
can write
the above expression as
\be
 {\cal I}m \Pi(p^2) = \frac{-m^2}{\sqrt{s}} \left (\frac{s-s_{th}}{m^2-s_{th}}
\right)^{1/2} \Gamma \; \theta(s-s_{th}).
\ee

The influence of the kaon width in the propagator is only felt
for $\sqrt{s}$ values near the kaon mass. Therefore the propagator can be
greatly
simplified by neglecting all but the first term in an expansion of ${\cal
I} m \Pi(s)$ around
$s =m^2$. More precisely,
\be
 {\cal I} m \Pi(s) = - m \Gamma \; \left ( 1 + {\cal O} \left( \frac{x \;
\Gamma}{m-s_{th}} \right) \right)
\ee
for $m-x\Gamma
\leq \sqrt{s} \leq m+x\Gamma$, with $x$ an arbitrary number such that
$x\Gamma/m
<<1$ and with $s_{th}<<m$.

Since $\Gamma_S/(m_S -2m_{\pi}) \sim O(10^{-14})$ \cite{pdg}, the
form of the propagator with a constant width
\be
\frac{i}{p^2 -m^2 +im\Gamma \; \theta(p^2-p^2_{th})}\ \ ,
\ee
turns out to be an extremely good approximation for the renormalized
propagator.

 Therefore, the space-time representation of the spin zero propagator for
the unstable particle can be written as
\be
\Delta_R (x'-x) \ = \ i \; \int \ \frac{d^4 p}{(2\pi)^4} \
\frac{e^{-ip.(x'-x)}}{p^2-m^2+im\Gamma \; \theta(p^2-p^2_{th})}.
\ee

Similarly as done above for the stable particle propagator, we would like to
express explicitly the time dependence of $\Delta_R(x'-x)$. It becomes
convenient to separate the propagator into two pieces :
\be
\Delta_R(x'-x)=\Delta_R^{(1)}(x'-x)+\Delta_R^{(2)}(x'-x)
\ee
with
\ba
\Delta_R^{(1)}(x'-x)&=& i \; \int \ \frac{d^4 p}{(2\pi)^4} \
\frac{e^{-ip.(x'-x)}}{p^2-m^2+im\Gamma} \nonumber \\ \\
\Delta_R^{(2)}(x'-x) &=& i \; \int \frac{d^3p}{(2\pi)^3}
\int^{p^0_{th}}_{-p^0_{th}} \frac{dp^0}{2\pi} e^{-ip.(x'-x)} \left\{
\frac{1}{p^2-m^2}-\frac{1}{p^2-m^2+im\Gamma} \right\} \nonumber
\ea
where $p^0_{th}=\sqrt{\vec{p}^{\ 2}+p^2_{th}}$.

   Using the condition $\Gamma/(m-\sqrt{p^2_{th}})<<1$,
we can show that \[
\Delta_R^{(2)}(x'-x) \sim {\cal O}\left(\frac{ \Gamma}{m-\sqrt{p^2_{th}}}
\right), \]
which allows to write
\be
\Delta_R(x'-x)= i \; \int \ \frac{d^4 p}{(2\pi)^4} \
\frac{e^{-ip(x'-x)}}{p^2-m^2+im\Gamma} \left( 1 + {\cal O} \left(
\frac{\Gamma}{m-\sqrt{p^2_{th}}} \right) \right).
\ee

In order to made explicit the time dependence of the unstable
propagator let us use the following pole decomposition
\be
p^2 - m^2 + i \Gamma m \ = \ \Biggl(p_0 - E + \frac{i\Gamma m}{2E}\Biggr)
\Biggl(p_0 +
E - \frac{i
\Gamma m}{2E}\Biggr) \Biggl(1+{\cal O}\Biggl(\frac{\Gamma^2}{m^2}
\Biggr)\Biggr)
\ee
where $E = \sqrt{\vec{p}^{\; 2} + m^2}$.

Therefore, by neglecting very small terms of order $10^{-14}$, the
contour integral in the complex $p^0$ plane with the poles located at
$\pm (E-im\Gamma/2E)$ gives
\ba
\Delta_R (x'-x) &=&
\int \ \frac{d^3 p}{(2\pi)^3} \ \frac{
e^{i\vec p. (\vec x' - \vec x \,)}
e^{-iE(t'-t)}
}{2E} \
e^{
- \frac{\Gamma}{2}
\frac{m}{E} (t'-t)} \theta (t'-t)
\nonumber \\
& &
+ \; \int \ \frac{d^3 p}{(2\pi)^3} \frac{e^{i \vec p.(\vec x - \vec x' \,)}
e^{-iE(t-t')}}{2E} e^{-\frac{\Gamma}{2}\frac{m}{E}(t-t')} \theta (t-t').
\ea
The interpretation is similar to the one for the stable particle, except
for the
decay constant $\Gamma$ which expresses the unstability of the particle and
antiparticle.
The case of $K^0 \overline{K^0}$ system considered in this paper is
more involved,  because the propagator is a 2$\times$2 matrix. This
problem is circumvented by performing the diagonalization (see Eq. (8))
 before doing the contour integration.

  Notice that $\tau=t'-t$ is the time elapsed between the production and
decay
locations of the resonance. Note also that, contrary to non-relativistic
approaches, the factor $m/E$ naturally appears in the exponential decay
factor. Therefore, no boost transformations are required to relate the {\it
proper} time to the time parameter of a moving particle. Of course, the
exponential decay takes its usual form $e^{-\Gamma \tau/2}$ \cite{kabir} in
the rest frame of the resonance.

 \section{Applications}

  In this section we compute the full S-matrix amplitudes for the
production and decay of neutral kaons as studied in CPLEAR and DAPHNE
experiments. Then, we derive the time evolution of these transition
amplitudes and introduce the CP violation parameters intrinsic to our
description.

\subsection{CPLEAR experiment}
At the CPLEAR experiment \cite{cplear}, $K^0$ and $\overline{K^0}$ are
produced at point $x$ in the strong interaction annihilation of $p\bar{p}$,
and subsequently decay at point $x'$ to $\pi^+\pi^-$ by the effects of weak
interactions. The production
mechanisms of $K^0$ and $ \overline{K^0}$ are $p\bar p\rightarrow K^0 K^-
\pi^+,\
\ \overline{K^0} K^+ \pi^-$, thus neutral kaons can be tagged by identifying
the accompanying charged kaon \cite{cplear}. After their production, $K^0$ (or
$\overline{K^0}$) oscillates between its two components $K_L$ and $K_S$
before decaying to the 2$\pi$ final states. We are
interested in the description of the time evolution of the full decay
amplitude and its interference phenomena. It is interesting to note that
despite the fact that charged kaons and pions have similar lifetimes as
$K_L$, they can be treated as asymptotic particles in the present case.

In order to relate the different S-matrix amplitudes, let us
first consider the production mechanism of $K^0 \overline{K^0}$. Since
strong interactions conserve strangeness, we have
\be
{\cal M} (p \overline p \to K^- \pi^+ \overline{K^0}) \ = \ {\cal M} (p
\overline p
\to K^+ \pi^- K^0) = 0,
\ee
which, according to equation (3), implies
$$
 {\cal M} (p \overline p \to K^- \pi^+ K_1) \ = \ {\cal M} (p \overline p
\to K^- \pi^+  K_2) \ \equiv \ A
\eqno{(26a)}
$$
$$
{\cal M} (p \overline p \to K^+ \pi^- K_1) \ = \ -{\cal M} (p \overline p
\to K^+
\pi^- K_2) \ \equiv \ B.
\eqno{(26b)}
$$
Assuming CPT invariance we obtain
$$
{\cal M} (p \overline p \to K^- \pi^+ K^0) \ = \ {\cal M} (p \overline p
\to K^+
\pi^- \overline{K^0}) \ \equiv \ C.
\eqno{(26c)}
$$
Collecting all these constraints, we get
$$
A \ = \ B \ = \ \frac{C}{\sqrt{2}}.
\eqno{(26d)}
$$
\setcounter{equation}{26}
Now, let us first consider the complete process for the production of a
$K^0$ decaying into $\pi^+\pi^-$
 \be
p(q) + \overline p (q') \to K^- (k) + \pi^+ (k') + K^0 (p) \to
K^-(k) + \pi^+(k') + \pi^+ (p_1) + \pi^- (p_2) \ .
\ee
The full amplitude corresponding to this process can be written (the
subscript $+-$ refers to the charges of the two pions from $K^0$
decay):
\ba
T_{+-} =&&
\int d^4 x \; d^4 x' \;
e^{i(p_1+p_2).x'}
\left( {\cal M} (K_1 \to \pi^+\pi^-), \  {\cal M} (K_2 \to \pi^+ 
\pi^-)\right)
  \\
&& \times \Delta^{K_1K_2}_R (x'-x)\left(\begin{array}{c}
{\cal M} (K^0 \to K_1) \\
{\cal M} (K^0 \to K_2) \end{array} \right)
 {\cal M} (p \overline p \to K^- \pi^+ K^0) \; e^{i(k+k'-q-q').x} \nonumber
\ea
where $\Delta_R^{K_1K_2}(x'-x)$ is the propagator matrix for the coupled
$K_1-K_2$ system in configuration space.

 With the help of equations (8) and (26), this gives
\ba
T_{+-} &= &
\int d^4 x \; d^4 x' \; e^{i(p_1+p_2).x'}\left( {\cal M} (K_1 \to
\pi^+\pi^-), \
{\cal M} (K_2 \to \pi^+ \pi^-)\right) \nonumber\\
&&\times \; i \; \int \frac{d^4 p}{(2\pi)^4} \; e^{-ip.(x'-x)}
 \frac{1}{1-\hat \varepsilon^2}
\left(\begin{array}{ccc}
1&\hat \varepsilon \\
\hat \varepsilon&1
\end{array} \right)
\left(\begin{array}{ccc}
d^{-1}_S(p) & 0 \\
0 & d^{-1}_L (p)
\end{array} \right)
\left(\begin{array}{ccc}
1&-\hat \varepsilon \\
-\hat \varepsilon&1
\end{array} \right) \nonumber\\
&& \times \left(\begin{array}{c}
{\cal M} (K^0 \to K_1) \\
{\cal M} (K^0 \to K_2) \end{array} \right)
\sqrt{2} \; A \;
 e^{i(k+k'-q-q').x}
\ea

Inserting equations (3) and (7), we obtain
\ba
T_{+-} &= &
i \;\int d^4 x \; d^4 x' \; \frac{d^4 p}{(2\pi)^4} \; e^{i(p_1+p_2-p).x'}
\; e^{i(k+k'-q-q'+p).x} \;
\frac{A}{1+\hat \varepsilon}
\nonumber\\
&&
\times
\left \{[{\cal M} (K_1 \to \pi^+\pi^-)+\hat \varepsilon \; {\cal M} (K_2 \to
\pi^+\pi^-)] \; \frac{1}{p^2-m^2_S+im_S\Gamma_S} \right.
\nonumber\\
&& \ \ \ \ \left. +
[\hat \varepsilon \; {\cal M} (K_1 \to \pi^+ \pi^-) + {\cal M} (K_2 \to
\pi^+\pi^-)] \; \frac{1}{p^2-m^2_L +im_L\Gamma_L}\right \}.
\ea

An expression depending on the four-momentum is obtained by performing all
the integrals :
\ba
T_{+-} &= &
i \; (2\pi)^4 \; \delta^{(4)} (q+q'-k-k'-p_1-p_2) \frac{A}{1+\hat \varepsilon}
\; {\cal M} (K_1 \to \pi^+\pi^-) \nonumber\\
& &
\times \; \left\{\frac{1+\chi_{+-} \hat \varepsilon}{(p_1+p_2)^2 - m^2_S +
i m_S \Gamma_S}
+ \frac{\hat \varepsilon + \chi_{+-}}{(p_1+p_2)^2 - m^2_L + i m_L
\Gamma_L}\right\}
\ea
where
\be
\chi_{+-} \equiv \frac{{\cal M} (K_2 \to \pi^+\pi^-)}{{\cal M} (K_1 \to \pi^+
\pi^-)}\ \ ,
\ee

\par\noindent
is the parameter describing direct CP violation in our approach.

Another possibility is to keep the time dependence, proceeding as in
section 3, on the diagonalized propagator of Eq.(30). In that case, the
amplitude of an originally pure $K^0$ state decaying into $\pi^+ \pi^-$ 
reads 
\ba
T_{+-} &=&
(2\pi)^4 \; \delta^{(4)} (q+q'-k-k'-p_1-p_2) \; \frac{A}{1+\hat
\varepsilon} \; {\cal M} (K_1 \to \pi^+ \pi^-) \\
       & & \times \left \{(1+\chi_{+-} \hat \varepsilon) \int \frac{dt}{2 E_S}
\left[e^{-i(E_S-E)t} \; e^{-\frac{1}{2}\Gamma_S \frac{m_S}{E_S}t} \; \theta(t)
+ e^{i(E_S - E)t} \; e^{\frac{1}{2} \Gamma_S \frac{m_S}{E_S} t} \; \theta
(-t)\right] \right. \nonumber\\
       & & \ \ \ \ \ + \left. (\hat \varepsilon + \chi_{+-}) \int
\frac{dt}{2E_L}
\left[ e^{-i(E_L-E)t} \; e^{-\frac{1}{2} \Gamma_L \frac{m_L}{E_L}t} \;
\theta (t)
+ e^{i(E_L-E)t} \; e^{\frac{1}{2} \; \Gamma_L \frac{m_L}{E_L}t} \; \theta
(-t)\right] \right\} \nonumber
\ea
where $E=p^0_1+p^0_2$ is the total energy of the $\pi^+ \pi^-$ system and
$E_{S,L} = \sqrt{(\vec p_1 + \vec p_2)^2 + m^2_{S,L} }$.

The time $t$ in eq. (33) has been defined as the time elapsed from
the production to the decay locations of $K^0$.
 Thus, the transition amplitude ${\cal T} (t)$ describing the
time evolution of the system for $t>0$ is given by the integrand
proportional to $\theta(t)$ in eq. (33), namely
\ba
{\cal T_{+-}} (t) &=& (2\pi)^4 \; \delta^{(4)} (q+q'-k-k'-p_1-p_2) \;
\frac{A}{1+\hat \varepsilon} \;
{\cal M} (K_1 \to \pi^+\pi^-) \; e^{iEt} \;
\nonumber\\
&& \times \left \{\frac{1+\chi_{+-}\hat \varepsilon}{2E_S} \; e^{-iE_St} \;
e^{-\frac{1}{2} \Gamma_S \frac{m_S}{E_S} t}
 + \frac{\hat \varepsilon + \chi_{+-}}{2E_L} \; e^{-iE_Lt} \; e^{-\frac{1}{2}
\Gamma_L \frac{m_L}{E_L} t} \right \}\ . \ \ \ \ \ \ \ \ \ \
\ea

Let us now consider the analogous process where a pure $\overline{K^0}$
state is initially produced and then decay to $\pi^+\pi^-$, {\it i.e.}
$p\bar{p} \to K^+\pi^-\overline{K^0} \to K^+ \pi^- \pi^+\pi^-$. Following
the same procedure as in the case of $K^0$ production and decay, we compute
the following expression for the time evolution of $\overline{K^0}$
decays
\ba
\overline{{\cal T}_{+-}\ } (t) &=& (2\pi)^4 \; \delta^{(4)} (q+q'-k-k'-p_1-p_2)
\frac{A}{1-\hat \varepsilon} \; {\cal M} (K_1 \to \pi^+ \pi^-) \;
e^{iEt} \nonumber \\
&& \times \left \{\frac{1+\chi_{+-} \hat \varepsilon}{2E_S} \; e^{-iE_S t}
\; e^{-\frac{1}{2} \Gamma_S \frac{m_S}{E_S}t}
 - \frac{\hat \varepsilon + \chi_{+-}}{2E_L} \; e^{-iE_L t} \;
e^{-\frac{1}{2} \Gamma_L
\frac{m_L}{E_L} t}
\right \}\ . \ \ \ \ \ \ \  \ \ \ \
\ea
  Let us notice that if we were interested in the $\pi^0\pi^0$ decay mode
of neutral kaons, we would have to replace in Eqs. (34) and (35) ${\cal
M}(K_1 \rightarrow
\pi^+\pi^-)$ by ${\cal M}(K_1 \rightarrow \pi^0\pi^0)$ and $\chi_{+-}$ by
$\chi_{00}$ where
\be
\chi_{00} \equiv \frac{{\cal M} (K_2 \to \pi^0\pi^0)}{{\cal M} (K_1 \to \pi^0
\pi^0)}\ .
\ee

  Using Eqs. (34) and (35), we can express the measurable ratio of
CP-violating to
CP-conserving decay amplitudes of $K_L$, $K_S$ states in terms of the
CP-violating parameters proper to our approach:
\be
\eta^{+-} \equiv \frac{{\cal M}(K_L \to \pi^+\pi^-)}{{\cal M}(K_S \to
\pi^+ \pi^-)}
= \frac{\hat \varepsilon + \chi_{+-} }{1+\chi_{+-}\hat \varepsilon}\ ,
\ee
and
 \be
\eta^{00} \equiv  \frac{{\cal M}(K_L \to \pi^0\pi^0)}{{\cal M}(K_S \to
\pi^0 \pi^0)}
=\frac{\hat \varepsilon + \chi_{00} }{1+\chi_{00}\hat \varepsilon}\ \ .
\ee
As is well known, the parameters $\eta^{+-}$ and $\eta^{00}$ are commonly
used to express the violation of CP in the two pion
decays of $K_L$ (see for example pages 422-425 in \cite{pdg}). Note
that the above relations between measurable quantities and the parameters
that quantify direct and indirect violation of CP, are derived {\it without}
relying on assumptions
based on isospin symmetry, contrary to the relations obtained for the
$\eta$ parameters in terms of the usual parameters $\varepsilon$ and
$\varepsilon'$. Furthermore, it can be explicitly shown that the 
parameters $\eta^{+-}$ and $\eta^{00}$ are independent of
the phase convention chosen for $K^0,\ \overline{K^0}$, which is not 
the case for $\hat \varepsilon$ and $\chi_{+-,\ 00}$. 

Using the isospin symmetry and the Wu-Yang phase convention \cite{mrr}, we 
observe that $\epsilon=\hat \varepsilon$ (see Ref. \cite{pdg}, p.102) and 
the 
parameters $\chi_{+-,\ 00}$ are expected to be very small so that terms
of $O(\chi_{ij} \varepsilon)$ can be neglected in the above equations.
In that limit, we obtain 
\ba
\chi_{+-} &=& \epsilon' \nonumber \\
\chi_{00} &=& -2 \varepsilon' . \nonumber
\ea

Finally, let us mention that Eqs. (34) and (35) reduce to the current
expressions for the time evolution used in the analysis of the CPLEAR
collaboration \cite{cplear}, when we choose the center of mass frame
($ \vec p_1 + \vec p_2 = \vec 0 $) of
the two pion produced in $K^0-\overline{K^0}$ decays.

\subsection{Neutral kaon production at DAPHNE}

In this section we consider the oscillations of the pair of neutral kaons
produced in $e^+e^-$ annihilations at DAPHNE \cite{daphne}. The results
obtained in the present formalism for the $K^0- \overline{K^0}$ system can
be straightforwardly generalized to describe the same phenomena in pair
production of neutral $B$ mesons in the $\Upsilon(4s)$ region
\cite{slac}.

  Neutral and charged kaons will be copiously produced ( $\sim 10^9$
pairs $K^0\overline{K^0}$/year) in
$e^+e^-$ collisions operating at a center of mass energy around the mass
of the $\phi(1020)$ meson \cite{daphne}. The $\phi$ mesons produced in
$e^+e^-$
annihilations decay at point $x$ into $K^0\overline{K^0}$ pairs, and
subsequently each neutral kaon oscillates between its $K_L-K_S$
components before decaying to final states $f_1(p)$ and $f_2(p')$ at
spacetime points $y$ and $z$ :
\be
\phi(q) \to K^0 \overline{K^0} \to f_1(p) f_2(p')\
\ee
where $q,\ p$ and $p'$ are the corresponding four-momenta.

  Since each final state can be produced by either $K^0$ or
$\overline{K^0}$, we must add coherently the two amplitudes arising from
the exchange of $K^0$ and $\overline{K^0}$ as intermediate states.
Taking into account the charge conjugation properties of the electromagnetic
current, the pair of neutral kaons is found to be in an antisymmetric state
\cite{lipkin68}. Thus, the relative sign of the
two contributions to $\phi \to f_1f_2$ decays must be negative. The
S-matrix amplitude for the process indicated in Eq. (39) is
\ba
T_{f_1f_2} &=& \int d^4 x \; d^4 y \; d^4 z \; e^{ip.y+ip'.z}
{\cal M} (\phi \to K^0 \overline{K^0}) \; e^{-iq.x} \nonumber \\
& & \times \left[
\left\{
\left({\cal M} (K_1 \to f_1) , \
{\cal M} (K_2 \to f_1)\right) \Delta^{K_1K_2}_R (y-x)
\left(\begin{array}{c}
{\cal M} (K^0 \to K_1)\\
{\cal M} (K^0 \to K_2)
\end{array}\right) \right\}
\right. \nonumber \\
& & \ \ \ \ \ \times \left.
\left\{ \left({\cal M} (K_1 \to f_2) , \
{\cal M} (K_2 \to f_2)\right) \Delta^{K_1K_2}_R (z-x)
\left(\begin{array}{c}
{\cal M} (\overline{K^0} \to K_1) \\
{\cal M} (\overline{K^0} \to K_2)
\end{array} \right) \right\}
\right. \nonumber \\
& & \ \ \ \ \ - \left.
\left( \mbox{same expression with $K^0 \leftrightarrow \overline{K^0}$ }\right)
\right].
\ea
Let us define ${\cal M}_{ij} \equiv {\cal M} (K_i \to f_j)$.
With the help of Eqs. (3) and (8), we can reexpress the previous amplitude as
 \begin{eqnarray}
T_{f_1f_2} &=& \int d^4 x \; d^4 y \; d^4 z \;e^{ip.y+ip'.z-iq.x}
 \int{d^4 k \over (2 \pi)^4}{d^4 k' \over (2 \pi)^4} \;
e^{-ik.(y-x)-ik'.(z-x)} \;
\frac{ (-1)}{ 2(1-\hat \varepsilon^2)}
\nonumber \\
& & \times \Biggl\{
\biggl(
\frac{ {\cal M}_{11} + \hat \varepsilon {\cal M}_{21} }{d_S (k)}
+ \frac{ \hat \varepsilon {\cal M}_{11} + {\cal M}_{21} }{d_L (k)}
\biggr)
\biggl(
\frac{ {\cal M}_{12} + \hat \varepsilon {\cal M}_{22} }{d_S (k')}
- \frac{\hat \varepsilon {\cal M}_{12} + {\cal M}_{22} }{d_L (k')}
\biggr)  \hspace*{29mm}
\nonumber\\
& &
\phantom{\times \Biggl\{}
-\bigg(
\frac{ {\cal M}_{11} + \hat \varepsilon {\cal M}_{21} }{d_S (k)}
- \frac{\hat \varepsilon {\cal M}_{11} + {\cal M}_{21} }{d_L (k)}
\bigg)
\bigg(
\frac{ {\cal M}_{12} + \hat \varepsilon {\cal M}_{22} }{d_S (k')}
+ \frac{\hat \varepsilon {\cal M}_{12} + {\cal M}_{22}}{d_L (k')}
\bigg)
\Bigg\}
\nonumber\\
& &
\times \; {\cal M} (\phi \to K^0 \overline{K^0})
\hspace*{84mm}
\nonumber \\
&=& - (2\pi)^4 \; \delta^{(4)} (q-p-p')
\; {\cal M} (\phi \to K^0 \overline{K^0} ) \; \frac{1}{1-\hat \varepsilon^2}
\nonumber \\
& & \times \Bigg\{
-\frac{ {\cal M}_{11} + \hat \varepsilon {\cal M}_{21} }
{p^2-m^2_S+im_S\Gamma_S} \;
\frac{\hat \varepsilon {\cal M}_{12} + {\cal M}_{22}}{p'^2-m^2_L+im_L\Gamma_L}
\nonumber \\
& &
\phantom{\times \Bigg\{}
+ \frac{\hat \varepsilon {\cal M}_{11}+{\cal M}_{21}}{p^2-m^2_L+im_L\Gamma_L}
\;
\frac{ {\cal M}_{12} + \hat \varepsilon {\cal M}_{22} }{p'^2 - m^2_S +im_S
\Gamma_S}
\Bigg\}\ .
\end{eqnarray}

As in the previous subsection, the time evolution of the amplitude is obtained
through the insertion of the explicit time-dependent propagator into the
amplitude (41).
The result is
\begin{equation}
T_{f_1f_2} = \int dt \; dt' \; \biggl ({\cal T} (t,t') \: \theta (t) \:
\theta (t') + \
\mbox{other terms} \sim \theta (-t) \; \mbox{or} \; \theta (-t') \biggr)
\end{equation}
where $t$ and $t'$ are the times taken by unstable kaons to propagate
from the common production point ($x$) up to their disintegration into
$f_1$ at point $y$ and $f_2$ at point $z$, respectively.

  Thus, the explicit time evolution of the decaying amplitude is given by
\begin{eqnarray}
{\cal T} (t,t')& =& (2\pi)^4 \delta^{(4)}
(q-p-p') \; e^{ip^0t+ip^{0'}t'} \; {1 \over 4E_S E_L} \;
{\cal M} (\phi \to K^0 \overline{K^0}) \; \frac{1}{1-\hat \varepsilon^2} \\ 
&& \Biggl\{
-({\cal M}_{11} + \hat \varepsilon {\cal M}_{21})
(\hat \varepsilon {\cal M}_{12} + {\cal M}_{22})
\; e^{-iE_S (p) t-{1 \over 2} \Gamma_S {m_S \over E_S} t}
\; e^{-iE_L (p')t'-{1 \over 2} \Gamma_L {m_L \over E_L} t'}
\hspace*{30mm}
\nonumber\\
&& \ \ \ \ \ +(\hat \varepsilon {\cal M}_{11}+ {\cal M}_{21})
({\cal M}_{12} + \hat \varepsilon {\cal M}_{22})
\; e^{-iE_S (p') t'-{1 \over 2} \Gamma_S {m_S \over E_S} t'}
\; e^{-iE_L (p)t-{1 \over 2} \Gamma_L {m_L \over E_L} t} \Biggr\}
\nonumber
\end{eqnarray}
 where $E_{S,L} (p) \ \equiv \ \sqrt{\vec p^{\ 2} + m^2_{S,L}}$.

As we have already pointed out in the case of the CPLEAR experiment, no
boost transformations are required to adequate the time evolution of the
decay amplitude to a given reference frame. Observe that, due to the
initial antisymmetrisation of the $K^0  \overline{K^0}$ system,  ${\cal
T} (t,t) = 0$ if $f_1 = f_2$ and $p=p'$ as noted in Ref. \cite{lipkin68}.

\section{Discussion and Conclusions}

In this paper we have discussed a formalism based on quantum field theory 
which describes a system of two unstable coupled pseudoscalar particles. 
Properties related to the production and decay of these unstable states 
are consistently incorporated into the relativistic S-matrix amplitude 
which is the physically meaningfull object for a given process 
\cite{veltman}. 
Therefore, this formalism does not exhibit the limitations intrinsic to 
the Wigner-Weisskopf approximation or non-relativistic approaches that we 
have discussed in the introduction of this paper.

  We have applied this formalism to describe the time evolution of the 
$K^0$ and $\overline{K^0}$ decay amplitudes in CPLEAR and DAPHNE 
experiments. Since Lorentz covariance is implicit to the field theory 
approach, our results are valid in any reference frame contrary to the 
results obtained in other approaches which require boost transformations 
to relate the time parameters defined in rest and moving frames. Let us 
mention that other papers, which depart from the WW approximation, have 
appeared recently \cite{lipkin,kayser,lowe,srivastava}; however, they all 
present limitations mainly related to the introduction of a proper time 
parameter which is intrinsic to non-relativistic treatments. Notice also 
that an interference term showing time oscillations will appear in the 
decay probabilities obtained from Eqs. (34), (35) and (43). This term can 
be converted to a evolution in space by using the classical formula 
$t=(E/|\vec{p}|) |\vec{x}|$, which applies only to particles observed in 
the detector and not to (off-shell) unstable $K_L$ and $K_S$ states.

Finally, we would like also to stress that the present formalism allows 
to define direct ($\chi_{+-,\ 00}$) and indirect ($\hat \varepsilon$) 
CP-violating parameters for $K$ decays without relying on assumptions 
based on isospin 
symmetry. This is particularly suitable because the factorization of 
strong rescattering effects may not be fully justified for the study of 
CP violation in $B$ decays.

  Let us remark that the present formalism can be extended 
straightforwardly to treat the isospin violation in the pion 
electromagnetic form factor in the $\rho -\omega$ resonance region 
\cite{mowt}. 

\

{\bf Acknowledgements.} We thank Jean-Marc G\'erard et Jacques Weyers for 
useful discussions.

\end{document}